\documentclass[prd,twocolumn,a4paper,superscriptaddress,floatfix]{revtex4}
\usepackage{graphicx}
\usepackage{bbm}
\usepackage[all]{xy}
\usepackage{amsmath}
\usepackage{amssymb}
\usepackage{epstopdf}

\def\be{\begin{equation}}
\def\ee{\end{equation}}
\newcommand{\bq}{\begin{eqnarray}}
\newcommand{\eq}{\end{eqnarray}}
\newcommand{\bes}{\begin{subequations}}
\newcommand{\ees}{\end{subequations}}

\def\ben{\begin{eqnarray}}
\def\een{\end{eqnarray}}
\def\ba{\begin{array}}
\def\ea{\end{array}}

\begin{document}
\newcommand{\half}{{\textstyle\frac{1}{2}}}
\allowdisplaybreaks[3]
\def\a{\alpha}
\def\b{\beta}
\def\g{\gamma}\def\G{\Gamma}
\def\d{\delta}\def\D{\Delta}
\def\ep{\epsilon}
\def\et{\eta}
\def\z{\zeta}
\def\t{\theta}\def\T{\Theta}
\def\l{\lambda}\def\L{\Lambda}
\def\m{\mu}
\def\f{\phi}\def\F{\Phi}
\def\n{\nu}
\def\p{\psi}\def\P{\Psi}
\def\r{\rho}
\def\s{\sigma}\def\S{\Sigma}
\def\ta{\tau}
\def\x{\chi}
\def\o{\omega}\def\O{\Omega}
\def\k{\kappa}
\def\pa {\partial}
\def\ov{\over}
\def\br{\\}
\def\ud{\underline}

\newcommand\lsim{\mathrel{\rlap{\lower4pt\hbox{\hskip1pt$\sim$}}
    \raise1pt\hbox{$<$}}}
\newcommand\gsim{\mathrel{\rlap{\lower4pt\hbox{\hskip1pt$\sim$}}
    \raise1pt\hbox{$>$}}}
\newcommand\esim{\mathrel{\rlap{\raise2pt\hbox{\hskip0pt$\sim$}}
    \lower1pt\hbox{$-$}}}
\newcommand{\dpar}[2]{\frac{\partial #1}{\partial #2}}
\newcommand{\sdp}[2]{\frac{\partial ^2 #1}{\partial #2 ^2}}
\newcommand{\dtot}[2]{\frac{d #1}{d #2}}
\newcommand{\sdt}[2]{\frac{d ^2 #1}{d #2 ^2}}    

\title{Layzer-Irvine equation: new perspectives and the role of interacting dark energy}

\author{P.P. Avelino}
\email[Electronic address: ]{ppavelin@fc.up.pt}
\affiliation{Centro de Astrof\'{\i}sica da Universidade do Porto, Rua das Estrelas, 4150-762 Porto, Portugal}
\affiliation{Departamento de F\'{\i}sica e Astronomia, Faculdade de Ci\^encias, Universidade do Porto, Rua do Campo Alegre 687, 4169-007 Porto, Portugal}
\author{A. Barreira}
\email[Electronic address: ]{a.m.r.barreira@durham.ac.uk}
\affiliation{Departamento de F\'{\i}sica e Astronomia, Faculdade de Ci\^encias, Universidade do Porto, Rua do Campo Alegre 687, 4169-007 Porto, Portugal}
\affiliation{Centro de F\'{\i}sica do Porto, Rua do Campo Alegre 687, 4169-007 Porto, Portugal}
\affiliation{Institute for Computational Cosmology, Department of Physics, Durham University, South Road, Durham, DH1 3LE, U.K.}
\affiliation{Institute for Particle Physics Phenomenology, Department of Physics, Durham University, South Road, Durham, DH1 3LE, U.K.}

\begin{abstract}

We derive the Layzer-Irvine equation in the presence of a homogeneous (or quasi-homogeneous) dark energy component with an arbitrary equation of state. We extend the Layzer-Irvine equation to homogeneous and isotropic universes with an arbitrary number of dimensions and obtain the corresponding virial relation for sufficiently relaxed objects. We find analogous equations describing the dynamics of cosmic string loops and other p-branes of arbitrary dimensionality, discussing the corresponding relativistic and non-relativistic limits. Finally, we generalize the Layzer-Irvine equation to account for a non-minimal interaction between dark matter and dark energy, discussing its practical use as a signature of such an interaction.

\end{abstract} 
\maketitle

\section{Introduction}

The Newtonian energy conservation equation when generalized to an expanding cosmological background becomes
\be
{\dot E} + H(2K+U)=0,\label{eq:LI}
\ee
where a dot represents a total derivative with respect to physical time, $H$ is the Hubble parameter, $E=K+U$, with $K$ and $U$ being the peculiar kinetic and gravitational potential energies, respectively, of a system of non-relativistic particles interacting through gravity. This equation was derived independentely in \cite{1961PhDT.........2I,1963ApJ...138..174L} and it is known as the cosmic energy or Layzer-Irvine equation (see also \cite{citeulike:2083812}). Eq. (\ref{eq:LI}) is valid throughout the entire process of structure formation and in the ${\dot E}=0$ limit one recovers the usual virial relation $K=-U/2$ that holds for collapsed objects that have reached the state of hydrostatic equilibirum. 

The Layzer-Irvine equation has been establishing itself as one of the most renowned equations of modern cosmology with its many applications including the determination of the matter density, cluster mass and size, and the galaxy peculiar velocity field \cite{Davis:1997vg,citeulike:1284470,Fukugita:2004ee,Zaroubi:2004vu}. More recently, some authors \cite{Bertolami:2007zm,Bertolami:2011yp,Abdalla:2007rd,Abdalla:2009mt} have been using the Layzer-Irvine equation as a tool to detect a possible non-minimal interaction between the dark matter (DM) and the dark energy (DE) which, together, account for approximately 96$\%$ of the energy content of the Universe today \cite{Amanullah:2010vv,Komatsu:2010fb} and whose fundamental nature is still largely unkown. The existence of such an interaction would in general invalidate the energy balance dictated by Eq. (\ref{eq:LI}). Consequently, by measuring the properties of sufficiently relaxed structures, such as galaxy clusters, one may expect to be able to detect a signature of an interaction between dark matter and dark energy through deviations from the usual virial relation \cite{Bertolami:2007zm,Abdalla:2007rd,Abdalla:2009mt,Bertolami:2011yp}.

In this paper, our main goal is to present a broad discussion of the Layzer-Irvine equation in a generalized framework (see also  \cite{Shtanov:2010iy}). In Sec. \ref{sec:LIa} we start by deriving the Layzer-Irvine equation in the presence of a homogeneous dark energy component, extending it to Friedmann-Robertson-Walker (FRW) cosmologies with more than three spatial dimensions. In Sec. \ref{sec:PB} we show that the dynamics of cosmic string loops and other p-branes of arbitrary dimensionality are described by analogous equations, discussing the corresponding relativistic and non-relativistic limits. In Sec. \ref{sec:INT} the Layzer-Irvine equation is generalized to the case where the dark matter is non-minimally coupled to the dark energy background and the implications of such coupling are discussed. Finnally we conclude in Sec. \ref{conc}.


\section{Non-interacting homogeneous dark energy}\label{sec:LIa}

There is now overwhelming evidence for the Cosmological Principle which states that the Universe is homogeneous and isotropic on cosmological scales. According to Birkhoff's theorem, in the context of General Relativity, the gravitational field must vanish inside a spherical symmetric shell, which is in agreement with the Newtonian result. This allows for the use of Newtonian mechanics in the study of the evolution of matter density fluctuations on  scales much smaller than the Hubble radius. In this limit, the Lagragian for a system of point mass dark matter particles of mass $m_i$, whose trajectories are given by ${\bf r}_i=a(t) {\bf x}_i$ may be written as
\be
\mathcal{L}=\sum_i \left( \mathcal{K}_i-\mathcal{U}_i\right)\label{L1}\,,
\ee
where
\be
\mathcal{K}_i=\frac12 m_i {\dot {\bf r}}_i \cdot {\dot {\bf r}}_i=\frac12 m_i v_i^2 + \frac{d}{dt} \left(\frac12 m_i a {\dot a} x_i^2\right)-\frac12 m_i a {\ddot a} x_i^2 \label{K1}
\ee
is the kinetic energy associated with the mass $m_i$, $a$ is the scale factor, $t$ is the physical time, ${\bf x}_i$ are comoving coordinates, $x_i=|{\bf x}_i|$, ${\bf v}_i={\dot {\bf r}}_i-H {\bf r}_i$ is the peculiar velocity, $H={\dot a}/a$ is the Hubble parameter, $v_i=|{\bf v}_i|$ and 
\be
\mathcal{U}_i=-m_i\frac{G}{2}\sum_{j \neq i} \frac{m_j}{|{\bf r}_j-{\bf r}_i|}+\frac{2\pi G(1+3w)\rho_w}{3}m_i r_i^2\label{Phi}\,,
\ee
is the potential energy associated with the mass $m_i$. Here, $w=p_w/\rho_w$ is the equation of state of the dark energy which we assume to be homogeneous ($\rho_w$ and $p_w$ are respectively the dark energy density and pressure). We shall assume that the background evolution of the universe, given by $a(t)$, is fixed, depending only on the average values of the density and pressure of dark matter and dark energy. Consequently, in Eq. (\ref{L1}) it is sufficient to consider only the  inhomogeneous contribution to the gravitational potential energy due to DM-DM and DM-DE interactions. 

By performing the canonical transformation
\be
\mathcal{L} \to  \mathcal{L} - \frac{d}{dt} \left(\frac12 a {\dot a} \sum_i m_i x_i^2\right)\,,
\ee
the Lagrangian may be written as
\be
\mathcal{L} = K-U\,,
\ee
with
\bq
K&=&\frac12\sum_i \left(m_i v_i^2\right)\,,\label{Keq}\\
U&=& \sum_i U_i\label{Ueq}\,,
\eq
where
\be
U_i =\mathcal{U}_i + \frac12 a{\ddot a} m_i  x_i^2\,.\label{Varphi}
\ee
The first term on the right hand side of Eq. (\ref{Phi}) can be separated into two components: the contribution due to the homogeneous matter distribution plus the contribution due to the matter density perturbations. Then, using Eqs.  (\ref{Ueq}) and (\ref{Varphi}), one obtains
\bq
U&=&  -\frac{G}{2}\int \frac{\left[\rho_m({\bf r})-{\bar \rho}_m\right]  \left[\rho_m({\bf r}')-{\bar \rho}_m\right]}{|{\bf r}-{\bf r}'|} d^3 {\bf r} d^3 {\bf r}' +\nonumber\\
 &+&\left( \frac{2\pi G}{3} \left[\bar \rho_m+\rho_w\left(1+3w\right)\right] a^2 +  \frac{a{\ddot a}}{2}\right)  \sum_i \left( m_i x_i^2\right) = \nonumber\\
 &=&  -\frac{G}{2}\int \frac{\left[\rho_m({\bf r})-{\bar \rho}_m\right]  \left[\rho_m({\bf r}')-{\bar \rho}_m\right]}{|{\bf r}-{\bf r}'|} d^3 {\bf r} d^3 {\bf r}'\ \,, \label{U1eq}
\eq
where $\rho_m$ is the matter energy density, ${\bar \rho}_m$ is its average value and the last equality was obtained using the Raychaudhuri equation given by
\be
\frac{\ddot a}{a}+\frac{4\pi G}{3}\left({\bar \rho}_m+(1+3w)\rho_w\right)=0\,.
\ee

The Hamiltonian is given by
\be
\mathcal{H} = \sum_i \left(\frac{p_i^2}{2 m_i}-U_i\right)\,,
\ee
with $p_i=m_i v_i$ and the classical energy equation is
\be
\frac{d\mathcal{H}}{dt} = \frac{\partial \mathcal{H}}{\partial t} \label{eq:energy equation},
\ee
where the partial derivative with respect to time is computed at fixed particle comoving coordinates ${\bf x}_i$ and comoving momenta ${\bf p}_i/a=m_i {\dot {\bf x}}_i$. This way, one has  $U \propto a^{-1}$ and $K \propto a^{-2}$. Consequently, using Eq. (\ref{eq:energy equation}) one finally obtains
\be
{\dot E}+ H(2K+U)=0 \,.
\ee
This shows that the minimally-coupled  homogeneous dark energy does not explicitly enter the Layzer-Irvine equation. The effect of dark energy is felt only through the impact it has on the evolution of the Hubble parameter $H$. This generalizes the result in \cite{2010ARep...54..185C} (where $w=-1$) to any homogenous dark energy form. For relaxed objects with ${\dot E}=0$ one obtains the usual virial relation 
\be
K=-\frac{U}{2} \label{virial} \,.
\ee

\subsection{Extra dimensions}\label{sec:LIb}

It is interesting to generalize the above result to a $N+1$-dimensional FRW universe with $N>2$. In that case, at fixed particle comoving coordinates ${\bf x}_i$ and comoving momenta ${\bf p}_i/a=m_i {\dot {\bf x}}_i$, one has  $U \propto a^{-N+2}$ and $K \propto a^{-2}$, which leads to
\be
{\dot E}+ H\left(2K+(N-2)U\right)=0 \,.
\ee
Taking the case of sufficiently relaxed objects, for which ${\dot E}=0$ is a good approximation, then the virial relation becomes
\be
K=-\frac{(N-2)}{2}U \,,
\ee
which reduces to Eq. (\ref{virial}) if $N=3$.

\section{p-brane dynamics}\label{sec:PB}

The dynamics of maximally cosmic strings loops, domain walls as well as higher dimensional p-branes in a cosmological background has been studied in detail in \cite{Avelino:2007iq,Avelino:2008mv}. This work has recently been extended to account for the dynamics of cosmological p-brane networks \cite{Sousa:2011ew,Sousa:2011iu}.

\subsection{Cosmic Strings}\label{sec:PBa}

In the absence of non-gravitational interactions (as well as gravitational radiation backreaction) the evolution of the total energy $E$ of a cosmic string loop is given by \cite{ Avelino:2007iq}
\be
{\dot E}=2 H E \left(\frac12-{\bar v}^2\right)\,,\label{eq:string}
\ee
with
\bq
E&=&\mu a \int \gamma ds \,, \label{Edef}\\
{\bar v}^2 &=& \frac{\int v^2 \gamma ds}{\int \gamma ds}\,,
\eq
where $\mu$ is the energy per unit length, $ds$ is the infinitesimal comoving arclength, $v$ is the loop velocity at a particular point and  $\gamma=(1-v^2)^{-1/2}$. For very small loops (with $E/\mu \ll H^{-1}$)  it is in general a good approximation to consider that the expansion has, on average, no impact on the total energy. Hence, the average over a sufficiently long time of the total energy $\langle E \rangle_t$ and root-mean-square (RMS) velocity $\langle {\bar v}^2\rangle_t =1/2$ is approximately constant. 

In the case of a very large non-relativistic loop, the total energy can be decomposed into the potential energy associated with the loop length $U=\mu L \propto a$, and the kinetic energy associated to the loop motion  $K=\mu L {\bar v}^2/2 \propto a^{-3}$ (${\bar v} \propto a^{-2}$), where $L$ is the physical length of the loop. As a result, using Eq. (\ref{eq:energy equation}) one obtains,
\be
{\dot E}+ H\left(3K-U\right)=0 \,.\label{eq:string equation}
\ee
Eq. (\ref{eq:string equation}) is very similar to the Layzer-Irvine equation: in both equations the derivative with respect to physical time of the total energy is proportional to the Hubble parameter times specific linear combinations of the kinetic and potential energy terms. Note however that in the case of non-relativistic cosmic strings one cannot set $\dot E = 0$ and therefore there is no analogy with the gravitational virial relation. This happens because in the non-relativistic regime $K \ll U$ so that $E \sim U \propto a$. Another diference is that, contrary to Eq. (\ref{eq:string equation}) that has a relativistic version (Eq. (\ref{eq:string})), there is no relativistic generalization of the Layzer-Irvine equation.

\subsection{p-branes}\label{sec:PBb}

Analogously to the case of the Layzer-Irvine equation, we can also generalize the cosmic string case to higher dimensions. In the case of a p-brane, Eq. (\ref{eq:string}) generalizes to  \cite{Avelino:2008mv}
\be
{\dot E}=(p+1)HE \left(\frac{p}{p+1}-{\bar v}^2\right)\,.
\ee
with
\bq
E&=&\sigma_p a^p \int \gamma d \mathcal{A}\,, \label{Edefp}\\
{\bar v}^2 &=& \frac{\int v^2 \gamma d \mathcal{A}}{\int \gamma d \mathcal{A}}\label{vdefp}\,,
\eq
where $\sigma_p$ is the energy per unit p-dimensional area, $d \mathcal{A}$ is the infinitesimal comoving p-dimensional area, $v$ is the p-brane velocity at a particular point and  $\gamma=(1-v^2)^{-1/2}$. For very small p-branes (with $(E/\sigma_p)^{1/p} \ll H^{-1}$) the expansion has in general a very small impact on the time average of the total energy $\langle E \rangle_t$ and RMS velocity $\langle {\bar v}^2\rangle_t =p/(p+1)$ which are therefore roughly constant. On the other hand, for very large non-relativistic p-branes one has $U=\sigma_p A \propto a^p$ and  $K=\sigma_p A {\bar v}^2/2 \propto a^{-2-p}$ (${\bar v} \propto a^{-p-1}$), where $A=a^p \mathcal{A}$ is the physical p-dimensional area of the p-branes. As a result, the energy equation becomes
\be
{\dot E}+ H\left((2+p)K-pU\right)=0\,.
\ee
This equation generalizes Eq. (\ref{eq:string equation}) to p-branes of arbitrary dimension in N+1-dimensional homogeneous and isotropic FRW universes (with $p < N$). The similarities with the Layzer-Irvine equation are again very evident.

\section{Interacting dark energy}\label{sec:INT}

One of the ways to better understand the physics of dark energy is through its influence on the formation of large-scale structures in the Universe. In an accelerated Universe the characteristic timescale for linear perturbation growth may become large compared to the Hubble time. However, if dark energy and dark matter interact non-minimally, then dark energy influences the process of structure formation in a more active way, not only through its impact on the acceleration of the Universe.

The coupling between dark matter and dark energy adds new source terms to the usual Layzer-Irvine equation \cite{Bertolami:2007zm,Bertolami:2011yp,Abdalla:2007rd,Abdalla:2009mt} (Eq. (\ref{eq:LI})). These extra terms can be written, with all generality, as 
\be
\left.\frac{\partial K}{\partial t}\right|_{int}=\alpha(t)HK \,, \qquad \left.\frac{\partial U}{\partial t}\right|_{int}=\beta(t)HU\,,\label{int terms}
\ee
so that the generalized Layzer-Irvine equation becomes
\be
{\dot E} + H\left((2-\alpha)K+(1-\beta)U\right)=0 \,.\label{generalized LI}
\ee
The functions $\alpha(t)$ and $\beta(t)$ depend on the details of the process of energy and momentum transfer between dark matter and dark energy and are therefore model dependent \cite{He:2008tn}. For example, if dark energy decays into dark matter, then new particles with non-vanishing momentum may be continuously added to the system. This way it would be crucial for the computation of $\alpha(t)$ and $\beta(t)$ to know not only the rate of energy transfer but also the initial RMS velocities of the new particles. On the other hand, the coupling might also occur through the dependence of the mass of the dark matter particles on the value of the dark energy field (see, for example, \cite{Farrar:2003uw}). In this paper, the model dependence associated with different choices of coupling models is incorporated in the freedom to choose the evolution of the parameters $\alpha(t)$ and $\beta(t)$.

In \cite{Bertolami:2007zm,Bertolami:2011yp} the case with $\alpha=0$ was considered, with $\beta$ being related to the coupling strenght.  It was argued that $\beta$ could be determined by measuring the kinetic and potential energy of sufficiently relaxed structures such as galaxy clusters. The homogeneous dark energy case with $\alpha=\beta/2$ has also been considered in \cite{Abdalla:2009mt}. If $\alpha$ and $\beta$ are constants then the virial relation obtained assuming hydrostatic equilibrium (${\dot E}= 0$) is given by
\be
K = \frac{\beta-1}{2-\alpha}U\,.
\ee
We note however that in the presence of such an interaction one cannot, in general, assume hydrostatic equilibrium. This can only happen if $\alpha$ and $\beta$ are constant or, according to Eq. (\ref{generalized LI}), if their evolution is given by
\be
\beta(t)=(E+(1-\alpha(t))K)/U\label{last}\,,
\ee 
with constant $E$, which does not happen in general. These two cases are very special and consequently, in the presence of an interaction between dark matter and dark energy, gravitationally bound systems are not expected to reach virial equilibrium. As a result, deviation from the usual virial relation in galaxy clusters is therefore a general signature of a non-minimal coupling between dark matter and dark energy.

\section{Conclusions \label{conc}}

In this paper we studied the Layzer-Irvine equation and discussed some of its generalizations. In particular, we derived the Layzer-Irvine equation in the presence of a general homogeneous dark energy background showing that the final form of the equation is not affected explicitly by the dark energy component. We further generalized the equation and the virial relation to FRW  cosmologies with $N+1$ dimensions (with $N>2$). We have also demonstrated that the macroscopic dynamical energy equations of cosmic string loops and other p-branes of arbitrary dimensionality are, in the non-relativistic limit, analogous to the Layzer-Irvine equation. Finally, we generalized the Layzer-Irvine equation to account for a non-minimal interaction between dark matter and a homogeneous dark energy form. We have shown that, in general, gravitationally bound systems are not expected to reach hydrostatic equilibrium in the presence of a coupling between these two components. This constrasts with the usual assumption in the literature where the equilibrium relation ${\dot E}=0$ is assumed a priori. Hence, a non-minimal coupling between dark matter and dark energy will generally lead to the breakdown of the usual virial relation $K = -U/2$, providing a crucial signature of such an interaction.

\begin{acknowledgments}

This work is partially supported by FCT-Portugal through project CERN/FP/116358/2010.

\end{acknowledgments}


\bibliography{LI}

\end{document}